\documentclass[reprint]{revtex4-1}

\usepackage[utf8]{inputenc}
\usepackage{amsmath}
\usepackage{amssymb}
\usepackage{amsfonts}
\usepackage{graphicx}
\usepackage{siunitx}					% mise en forme des unités
\usepackage{color}					% gestion de différentes couleurs

\newcommand{\eff}{\mathrm{eff}}
\newcommand{\ext}{{\mathrm{ext}}}
\newcommand{\fluid}{{\mathrm{fluid}}}

\renewcommand{\d}{\mathrm{d}}

\begin{document}

\title{Resonance frequency shift of strongly heated micro-cantilevers} %Title of paper

\author{Felipe Aguilar Sandoval}
\affiliation{Universit\'e de Lyon \& CNRS, Laboratoire de Physique ENS Lyon (France)}
\affiliation{Present address: Universidad de Santiago de Chile, Departamento de F\'isica (Santiago, Chile)}

\author{Mickael Geitner}
\affiliation{Universit\'e de Lyon \& CNRS, Laboratoire de Physique ENS Lyon (France)}

\author{\'Eric Bertin}
\affiliation{Universit\'e Joseph Fourier \& CNRS, Laboratoire interdisciplinaire de Physique (Grenoble, France)}

\author{Ludovic Bellon}
\email{ludovic.bellon@ens-lyon.fr}
\affiliation{Universit\'e de Lyon \& CNRS, Laboratoire de Physique ENS Lyon (France) - Ludovic.Bellon@ens-lyon.fr}

\date{\today}

\begin{abstract}

In optical detection setups to measure the deflection of micro-cantilevers, part of the sensing light is absorbed, heating the mechanical probe. We present experimental evidences of a frequency shift of the resonant modes of a cantilever when the light power of the optical measurement set-up is increased. This frequency shift is a signature of the temperature rise, and presents a dependence on the mode number. An analytical model is derived to take into account the temperature profile along the cantilever, it shows that the frequency shifts are given by an average of the profile weighted by the local curvature for each resonant mode. We apply this framework to measurements in vacuum and demonstrate that huge temperatures can be reached with moderate light intensities: a thousand $\SI{}{\degree C}$ with little more than $\SI{10}{mW}$. We finally present some insight into the physical phenomena when the cantilever is in air instead of vacuum.

\end{abstract}

\maketitle 

\section{Introduction}

Silicon micro-cantilevers have today applications in many areas of measurement, as chemical and biological sensors~\cite{Lavrik-2004}, mass detectors~\cite{Thundat-1994,Dohn-2005}, flow meters~\cite{Barth-2005,Salort-2012}, or force sensors~\cite{Mamin-2001}. Thanks to their industrial production with tight tolerances, and the ability to integrate a sharp tip in the fabrication process, they are for example ubiquitous in atomic force microscopy (AFM)~\cite{Binnig-1986,Garcia-2002,Butt-2005,Garcia-2012}. The detection of their deflection is usually performed optically: optical lever scheme~\cite{Meyer-1988} in most of commercial systems, interferometric measurements in the most precise instruments~\cite{Rugar-1989,Schonenberger-1989,Mulhern-1991,Mamin-2001,Hoogenboom-2008,Paolino-2009-JAP,Jourdan-2009,Laurent-2012,Laurent-2013,Paolino-2013}. The former approach is very sensitive~\cite{Fukuma-2005} while remaining simple to set up, and the latter reaches the highest resolution in challenging experiments. The limit to the sensitivity of these optical systems is due to the shot-noise of the photodetectors: this white noise sets the lower limit of detectable deflections~\cite{Gustafsson-1994,Fukuma-2005}. Its level is linked to the intensity of light on the photodetectors: higher intensities translates into better signal to noise ratio. Using high power light sources is thus appealing.

The light focused on the micro-cantilever may however have some influence on its mechanical response: absorption of the radiation heats locally the cantilever~\cite{Marti-1992,Allegrini-1992,Thundat-1994,McCarthy-2005,Yang-2009,Milner-2010,Chen-2011}. Very high temperatures (several hundred of degrees) have been measured for example by Raman scattering for laser powers of the order of $\SI{10}{mW}$~\cite{McCarthy-2005,Milner-2010,Chen-2011}. Permanent surface modification~\cite{Kumar-2012} or unwanted damage to the tip \cite{Marti-1992} have been observed in some experiments, pointing at the temperature reaching the melting point of the material of the cantilever, above $\SI{1000}{\degree C}$. The effect will be more pronounced in vacuum noteworthily, since heat has no other efficient way out than through the tiny cross-section of the cantilever, producing a noticeable temperature rise of the probe. This photo-thermal effect can be useful in some applications, for example to drive a cantilever at its resonance in dynamic measurements~\cite{Marti-1992,Allegrini-1992,Ramos-2006,Kiracofe-2011,Bircher-2013}, or to heat the AFM tip and control local phase changes of a sample~\cite{Hamann-2004}. On the contrary, this can be an issue for temperature sensitive samples, or for low temperature experiments~\cite{Mamin-2001,Laurent-2012}, those aiming at reducing thermal noise for example.

Silicon is a poor mirror for visible light, with only $\SI{37}{\%}$ of reflectivity, thus radiation absorption is large in raw silicon micro-cantilevers~\cite{Green-2008}. Reflex coatings (Aluminium, Platinum, Gold) are often used to minimize this issue, but they add some viscoelastic damping to the probe mechanical response, lowering its quality factor and thus its performance in dynamic operation~\cite{Sandberg-2005,Paolino-2009-Nanotech,Li-2012,Li-2014}. Understanding the effect of heating a cantilever through the measurement system is thus important for its operation~\cite{Gysin-2004}. In this article, we present an approach to quantify the temperature rise and its consequences on the mechanical response of micro-cantilevers, in vacuum and in air. The analysis applies consistently across all flexural modes of the cantilever, a welcomed point in nowadays growing scope of application of multifrequency AFM~\cite{Garcia-2012}.

The article is organized as follows: in part \ref{section:exp}, we present experimental evidence of a frequency shift of the resonant modes of a cantilever when the light power of the optical measurement set-up is increased. This frequency shift is a signature of the temperature rise~\cite{Mertens-2003,Gysin-2004,Yang-2009,Milner-2010}, but its dependence on the mode number demonstrates that this temperature cannot be uniform on the cantilever. In part \ref{section:model}, we present an analytical model to take into account the temperature profile along the cantilever, and show that the frequency shifts are given by an average of the profile weighted by the local curvature for each resonant mode. In part \ref{section:discussion}, we apply this framework to measurements in vacuum. We demonstrate that huge temperatures can be reached with moderate light intensities: a thousand $\SI{}{\degree C}$ with little more than $\SI{10}{mW}$. In part \ref{section:air}, we finally present some insight into the physical phenomena at play when the cantilever is in air instead of vacuum, before giving a general conclusion to this work.

\section{Experimental data: frequency shift\label{section:exp}} 

\begin{figure}[htbp]
\begin{center}
\includegraphics{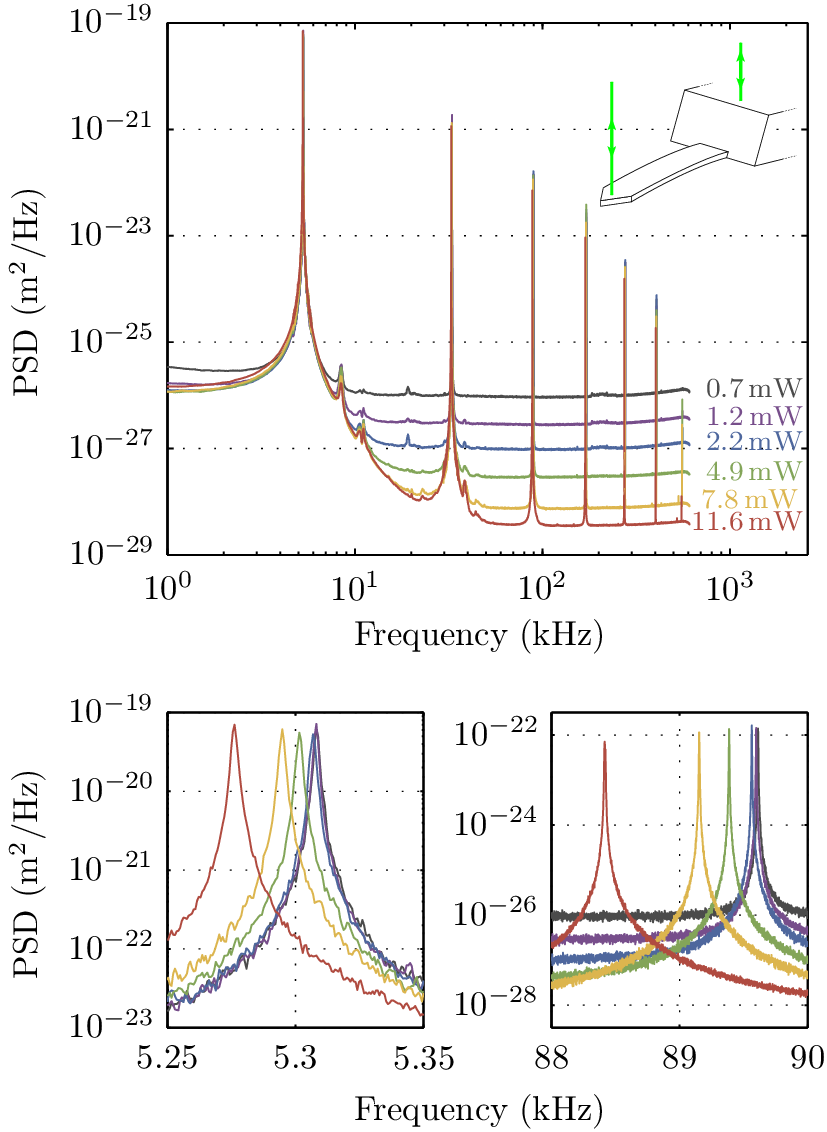}
\caption{(Color online) --- Thermal noise measurement on cantilever C100: we plot the power spectrum density (PSD) of thermal noise induced deflection as a function of frequency, for various incident light power $I$ on the lever ($I=\SI{0.7}{mW}$ to $\SI{11.6}{mW}$), in vacuum ($P=\SI{4}{Pa}$). The flat base line of the measurement is due to the shot noise on the photodetectors, it decreases when $I$ increases. All the resonance frequencies of the cantilever decrease when $I$ grows, as illustrated in the two bottom figures for modes 1 and 3. The inset of the top figure illustrates the principle of the measurement: the fluctuations of deflection are recorded through the interference of two laser beams, one reflected on the cantilever free end, the other on the chip holding the cantilever~\cite{Paolino-2013}.}
\label{spectra_variousI}
\end{center}
\end{figure}

We perform the experiments on two different geometries of cantilevers: cantilever C100 is $L=\SI{500}{\micro m}$ long, $W=\SI{95}{\micro m}$ wide, and $H=\SI{0.75}{\micro m}$ thick (NanoWorld Arrow TL8), while cantilever C30 is $L=\SI{500}{\micro m}$ long, $W=\SI{30}{\micro m}$ wide, and $H=\SI{2.8}{\micro m}$ thick (BudgetSensors AIO-TL). Both are uncoated tipless AFM silicon cantilevers. Geometrical dimensions were measured using a scanning electron microscope (SEM), with uncertainties around 1\% for $L$ and $W$, and 5\% for $H$. The thermal noise measurements are performed using a quadrature phase differential interferometer~\cite{Paolino-2013}, with the sensing laser beam focused on the free end of the cantilever, and the reference beam on the silicon chip where the cantilever is clamped (see inset of figure \ref{spectra_variousI}). We use a stabilized solid state laser from Spectra Physics, with a $\SI{100}{mW}$ maximum output power at $\SI{532}{nm}$. We couple this laser through an optical fiber into our measurement set-up, and an intensity $I$ up to $\SI{25}{mW}$ can be used for the sensing beam. This intensity can be tuned to lower levels using neutral density filters at the laser output, ensuring highly stable light intensity and frequency during the whole measurement process. 

We record the power spectral density (PSD) of the brownian deflection of the cantilever using our interferometer, and plot the results in figure~\ref{spectra_variousI}. Such a spectrum is usually computed by an average on 20 PSD, each evaluated on a $\SI{5}{s}$ deflection measurement sampled at $\SI{1.4}{MHz}$. The 6 first normal modes of the cantilever C100 are clearly seen of this spectra. A few higher order modes could also be measured, but we limited our study to these 6 modes. To ensure precise positioning of the sensing laser beam on the cantilever and avoid any spurious effect from thermal drifts, we actually use the 7th resonant mode of this cantilever : we look for the last spatial node of this mode close to the free end of the cantilever. Killing any thermal noise in the spectrum at the corresponding resonance frequency defines the longitudinal position of the sensing beam with better than $\SI{3}{\micro m}$ repeatability. Centering the laser spot on the cantilever is also done with the same precision killing any thermal noise corresponding to torsional modes of the lever. A similar procedure can be applied to cantilever C30, for which we characterize up to 4 modes.

The background noise of the instrument is as low as $\SI{5e-29}{m^{2}/Hz}$, as illustrated by the noise floor at the highest light intensity measurement in figure~\ref{spectra_variousI}. The main contribution to this background noise is the shot noise of the photodiodes used in our optical setup, it thus decreases when light intensity increases. It is therefore tempting to use the highest light power available to increase the dynamic of our measurement. We were able to lower the background noise down to $\SI{e-30}{m^{2}/Hz}$ for some cantilevers with a reflective coating. However, the sensing light is also partially absorbed by the cantilever, especially for raw silicon cantilevers presented in this article, leading to a heating of the sensor, potentially modifying its response or integrity. And indeed, a closer look on the resonant peaks of the PSDs in figure~\ref{spectra_variousI} demonstrate a light power dependence: the resonance frequencies shift when $I$ changes~\cite{Gysin-2004,Yang-2009,Milner-2010}. To study this effect, we fit each of these resonances with a lorentzian curve to extract the resonance frequency $f_{n}$ for each mode number $n$.

\begin{figure}[htbp]
\begin{center}
\includegraphics{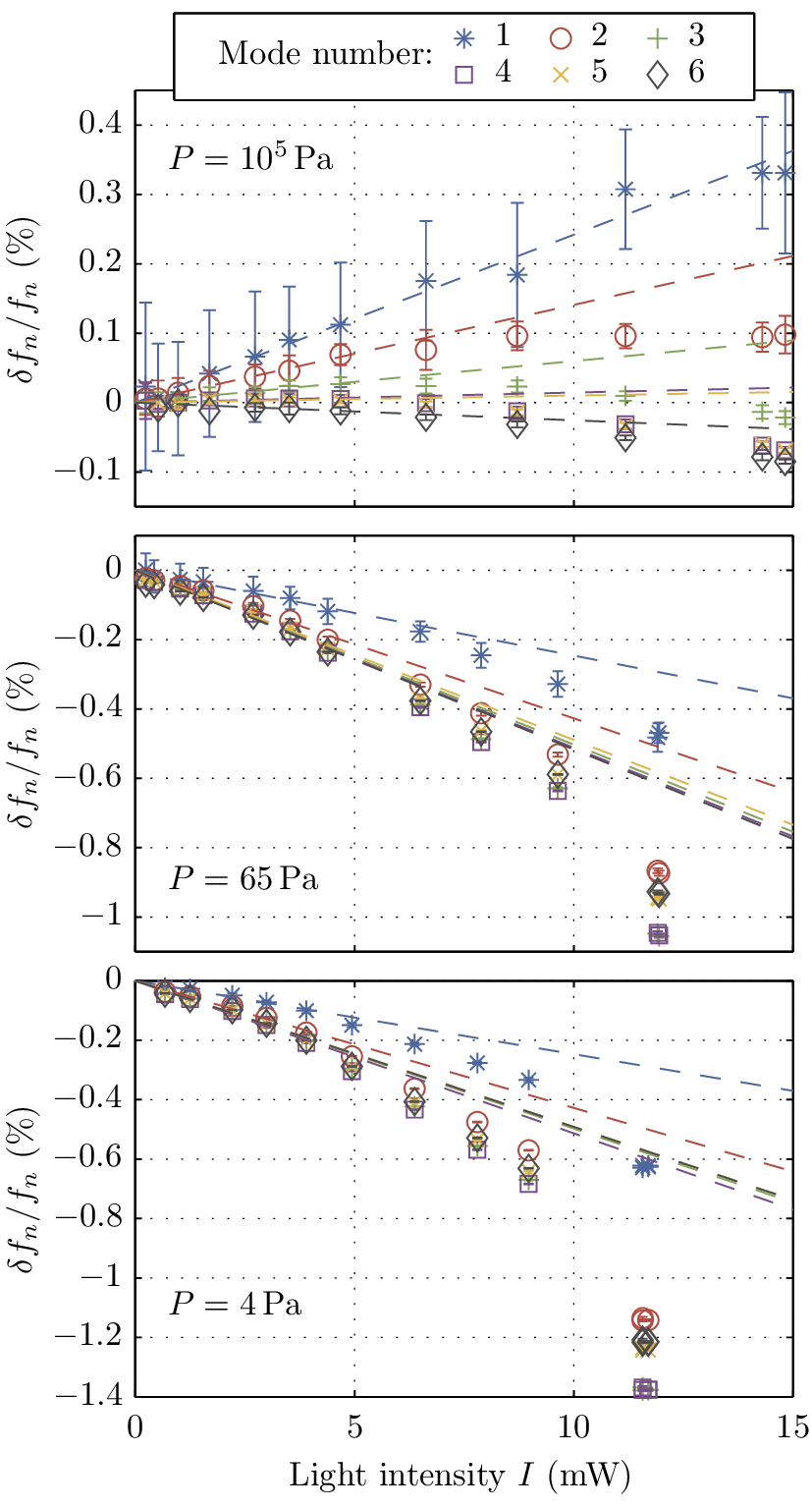}
\caption{(Color online) --- Cantilever C100: relative frequency shift $\delta f_{n}/f_{n}$ for modes $n=1$ to $6$ as a function of the incident light power $I$ on the lever, at atmospheric pressure $P=\SI{e5}{Pa}$, top) and in vacuum ($P=\SI{65}{Pa}$ middle, $P=\SI{4}{Pa}$ bottom). The dashed lines are linear fits on the initial part of the curves ($I<\SI{6}{mW}$). The curves shapes are nonlinear, and depends strongly on pressure and mode number.}
\label{fig:dfC100}
\end{center}
\end{figure}

\begin{figure}[htbp]
\begin{center}
\includegraphics{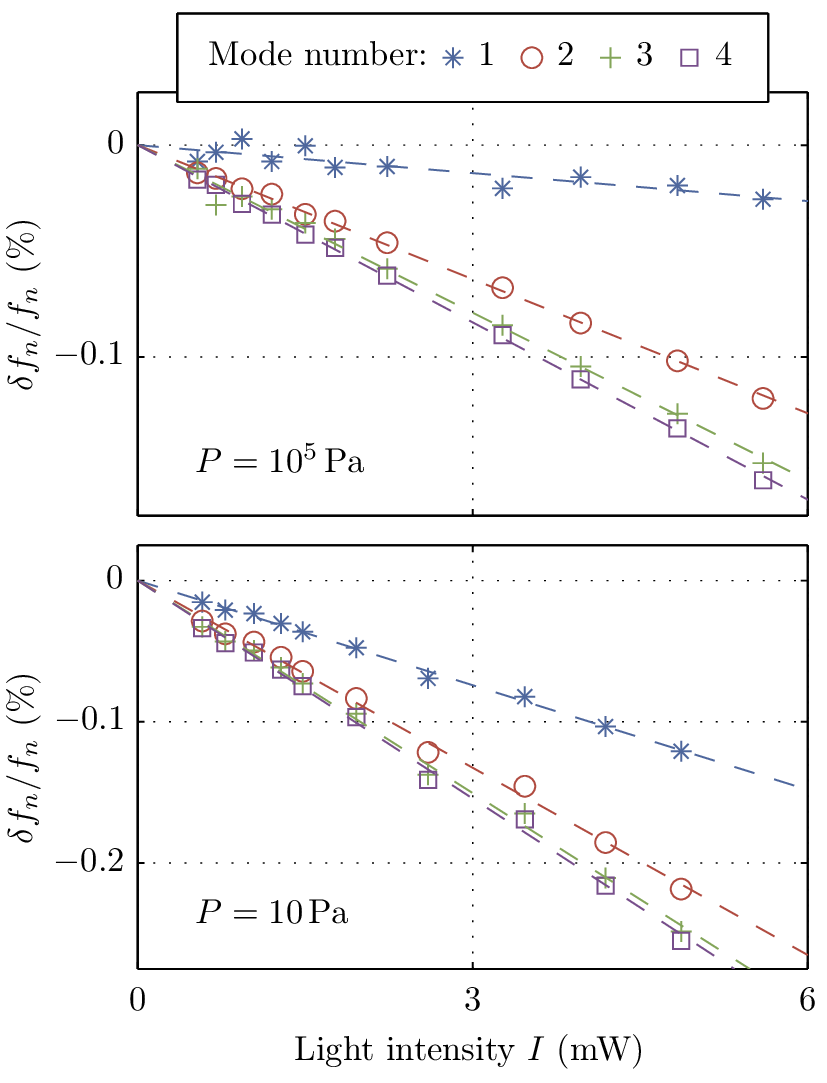}
\caption{(Color online) --- Cantilever C30: relative frequency shift $\delta f_{n}/f_{n}$ versus incident light power $I$ on the lever, at atmospheric pressure $P=\SI{1e5}{Pa}$, top) and in vacuum ($P=\SI{10}{Pa}$ bottom). The dashed lines are linear fits.}
\label{fig:dfC30}
\end{center}
\end{figure}

We plot in figures \ref{fig:dfC100} and \ref{fig:dfC30} the measured relative frequency shift $\delta f_{n}/f_{n}^{0}$ for each mode number $n$ as a function of the light intensity $I$ of the sensing laser beam at the cantilever end. $f_{n}^{0}$ is defined as the ordinate at the origin of a linear fit of $f_{n}$ versus $I$ for low laser powers, such that $\delta f_{n}=0$ for $I=0$. These curves depend on mode number, on the pressure of air around the cantilever, and on the cantilever shape. We also observe some nonlinear behaviors of the frequency shift as a function of light intensities for the larger cantilever C100 when $I>\SI{6}{mW}$. The laser spot size is tuned around $\SI{10}{\micro m}$ in diameter in these measurements to avoid most of light spilling out of the cantilever. The non-linear effects tend to occur sooner for smaller spot sizes.

A simple argument can be used to understand the global behavior of the resonance frequency~\cite{Mertens-2003,Gysin-2004,Yang-2009,Milner-2010}: each mode can be pictured as a harmonic oscillator of mass $m_{\eff}$, whose stiffness $k_{n}$ is proportional to the Young's modulus $E$ of silicon (the cantilever material). The increase of temperature, due to light absorption, induces a softening of the cantilever: the temperature coefficient of $E$ is negative~\cite{Walsh-1991,Bourgeois-1997,Li-2003,Gysin-2004,Hopcroft-2010,Masolin-2013}
\begin{equation}
\alpha_{E} = \frac{1}{E} \frac{\d E}{\d T} \approx \SI{-64e-6}{K^{-1}}
\end{equation}
The resonance frequencies of the normal modes should therefore decrease as light intensity increases:
\begin{align}
2\pi f_{n} & = \sqrt{\frac{k_{n}}{m_{\eff}}} \propto E^{1/2} \\
\frac{\delta f_{n}}{f_{n}^{0}} & \approx \frac{1}{2} \frac{\delta E}{E} \approx \frac{1}{2} \alpha_{E} \frac{\d T}{\d I} I
\end{align}
However, this simple argument does not explain why the frequency shift is mode number dependent (and even less how the shift can be positive in air for cantilever C100). In the next part, we show that the mode number dependence can be explained by the coupling between the spatial mode shape of the normal modes and the temperature profile in the cantilever. 

\section{Analytical description: frequency shift of resonant modes\label{section:model}}

In the Euler-Bernoulli framework to describe the micrometer-sized mechanical beam, we assume that the cantilever length $L$ is much larger than its width $W$, which itself is much larger than its thickness $H$. The flexural modes of the cantilever are supposed to be only perpendicular to its length and uniform across its width. The deformations can thus be described by the deflection $d(x,t)$, with $t$ the time, and $x$ the spatial coordinate along the beam. The equation of motion for the cantilever reads~\cite{Butt-1995,Sader-1998}:
\begin{equation} \label{eq0}
\mu(x) \frac{\partial^{2}d}{\partial t^{2}} + \frac{\partial^{2}}{\partial x^{2}}\left[E(x)I(x) \frac{\partial^{2}d}{\partial x^{2}}\right] = f^{\fluid}(x,t)+f^{\ext}(x,t)
\end{equation}
where $\mu$ is the mass per unit length, $EI$ the bending stiffness ($E$ Young's modulus, $I$ the second moment of inertia) of the cantilever, $f^{\fluid}(x,t)$ the force per unit length due to the surrounding fluid (if any) acting on the cantilever, and $f^{\ext}(x,t)$ the external force.

We model the influence of the fluid according to Sader's model~\cite{Sader-1998}: $f^{\fluid}(x,t)$ is supposed to be linear in the local deflection $d(x,t)$. We therefore write in Fourier space: 
\begin{equation} \label{eq:dissip-force-fluid}
f^{\fluid}(x,\omega) = \mu_{f} \omega^{2} \Gamma (\omega,\nu) d (x,\omega) 
\end{equation}
where $\mu_{f}=\pi \rho W^{2}/4$ is the mass per unit length of a cylinder of diameter $W$ of fluid of density $\rho$, $\nu$ is the fluid kinematic viscosity, and $\Gamma$ is the complex hydrodynamic function of a rectangular cantilever~\cite{Sader-1998}. The real part $\Gamma_{r}$ of $\Gamma$ translates into inertial effects, while the imaginary part $\Gamma_{i}$ accounts for the viscous damping. In vacuum, dissipation occurs only through internal mechanisms (viscoelastic damping) in the cantilever~\cite{Saulson-1990, Paolino-2009-Nanotech}. It can be modeled using a complex Young's modulus, but for simplicity we shall mainly consider viscous dissipation in this paper.

Due to the heating of the cantilever by the sensing laser beam, its temperature is not uniform. In line with previous geometrical hypotheses, we will suppose that the temperature profile is uniform across the section of the cantilever and time independent, thus of the form
\begin{equation}
T(x) = T_{0}+\theta(x)
\end{equation}
where $T_{0}$ is the clamping base temperature (we make the hypothesis that the clamping base is at room temperature thanks to its macroscopic size. $\theta(x)$ is thus the temperature difference profile along the cantilever. This temperature profile will locally change the physical properties of the cantilever and surrounding atmosphere if not in vacuum. As a first approximation for the cantilever, we will only consider the variation of Young's modulus and dissipation with temperature, and neglect any change in the physical dimensions. Indeed, the coefficient of linear thermal expansion of silicon $\alpha_{l} \approx \SI{2.6e-6}{K^{-1}}$ is one order of magnitude smaller than the temperature coefficient of its Young's modulus $\alpha_{E}\approx \SI{-64e-6}{K^{-1}}$ (numerical values at room temperature~\cite{Bourgeois-1997}). Moreover, we will only consider first order correction for the Young's modulus: $E(T)=E_{0}(1+\alpha_{E} \theta)$. Note that we only need to assume that $\alpha_{E} \theta \ll 1$ (which is a light assumption since it means $\theta \ll 1/\alpha_{E} = \SI{15000}{K}$), but not that $\theta \ll T_0$. We therefore rewrite equation~(\ref{eq0}) in Fourier space in time to explicit the temperature-dependence of the bending stiffness:
\begin{align} \label{eq:beam}
\frac{k_{0}}{3}\frac{\partial^{2}}{\partial x^{2}}\left[(1+\alpha_{E}\theta(x))\frac{\partial^{2}d}{\partial x^{2}}\right] \nonumber & \\ 
- (m+m_{f}\Gamma (\omega,\nu)) \omega^{2} d & =  L f^{\ext}(x,\omega)
\end{align}
where $m$ and $k_{0}$ are the cantilever mass and static stiffness ($k_{0}$ is defined at room temperature $T_{0}$), $m_{f}=L \mu_{f}$, and $x$ is now normalized to the length $L$ ($x=1$ at the free end of the cantilever). In absence of external forcing (normal modes), the boundary conditions for Eq.~(\ref{eq:beam}) correspond to the classical clamped ($x=0$) and free end ($x=1$) conditions:
\begin{align}
d(x{=}0,\omega)& = 0& \left . \frac{\partial d(x,\omega)}{\partial x}\right |_{x=0}& = 0  \nonumber \\
\left . \frac{\partial^{2}d(x,\omega)}{\partial x^{2}}\right |_{x=1}& = 0& \left . \frac{\partial^{3}d(x,\omega)}{\partial x^{3}}\right |_{x=1}& = 0  \label{eq:boundary}
\end{align}

\subsection{Naive approach : uniform temperature}

Let us first study the case where $\theta(x)=0$. We introduce the operator $\mathcal{L}_{0}= \partial^{4}/\partial x^{4}$ in the space $\mathcal{D}$ of functions matching boundary conditions (\ref{eq:boundary}). This operator is self-adjoint on this space for the integral scalar product :
\begin{equation}
\phi \odot \psi = \int_{0}^{1} \phi(x) \psi(x) \d x  \rightarrow \mathcal{L}_{0} \phi \odot \psi = \phi \odot \mathcal{L}_{0} \psi
\end{equation}
The eigenvectors of operator $\mathcal{L}_{0}$ form an orthonormal basis of $\mathcal{D}$. They are the classic normal modes of the Euler-Bernouilli model of the free-clamped mechanical beam:
\begin{equation} \label{eq:normal modes}
\phi_{n}^{0}(x)=(\cos \alpha_{n}x -\cosh \alpha_{n}x) + B_n (\sin \alpha_{n}x - \sinh \alpha_{n}x)
\end{equation}
with
\begin{equation}
B_n = \frac{\cos \alpha_{n}+\cosh \alpha_{n}}{\sin \alpha_{n} + \sinh \alpha_{n}}
\end{equation}

and where $\alpha_{n}$ is the $n^{th}$ solution of equation
\begin{equation} \label{eq:Cn}
1+ \cos \alpha_{n}\cosh \alpha_{n} = 0
\end{equation}
which leads to $\alpha_{1}=1.875$, $\alpha_{2}=4.694$, \ldots, and $\alpha_{n}=(n-1/2)\pi$ for large $n$. The eigenvalues $\lambda_{n}^{0}$ of the operator $\mathcal{L}_{0}$ corresponding to each normal mode are simply $\lambda_{n}^{0}=\alpha_{n}^{4}$:  $\mathcal{L}_{0} \phi_{n}^{0} = \lambda_{n}^{0} \phi_{n}^{0} = \alpha_{n}^{4} \phi_{n}^{0}$.

We project equation~(\ref{eq:beam}) on this basis to get the evolution of the amplitude of each mode:
\begin{equation} \label{eq:amplitudemodeT0}
\left[ - m_{\eff} \omega^{2} + \frac{k_{0}}{3} \alpha_{n}^{4} + i \omega \gamma_{\eff}\right] d_{n}(\omega)= F^{\ext}_{n}(\omega)
\end{equation}
with $d_{n}= d \odot \phi_{n}^{0}$, $F^{\ext}_{n}= L f^{\ext} \odot \phi_{n}^{0}$, $\gamma_{\eff}=\omega m_{f}\Gamma_{i}(\omega,\nu)$ and $m_{\eff}=m+m_{f}\Gamma_{r}(\omega,\nu)$. $\Gamma (\omega,\nu)$ is usually slowly varying with $\omega$, and dissipation is low (large quality factor), thus for every resonance a simple viscous damping is a good approximation. The conclusion would be the same with a viscoelastic damping, though the damping coefficient $\gamma_{\eff}$ would present an explicit mode dependence. We therefore have a collection of independent quasi-harmonic oscillators, of effective mass $m_{\eff}$, stiffness $k_{n}=\alpha_{n}^{4} k_{0}/3$, and damping coefficient $\gamma_{\eff}$.

The resonant pulsations $\omega_{n}$ of the oscillators are linked to the spatial eigenvalues by the dispersion relation:
\begin{equation} \label{eq:dispertion}
m_{\eff} \omega_{n}^{2} = \frac{k_{0}}{3} \alpha_{n}^{4} = \frac{k_{0}}{3} \lambda_{n}^{0}
\end{equation}
If we now consider a uniform temperature rise $\theta = \Delta T$, assuming $m_{\eff}$ to be constant, we only need to change the value of $k_{0}$ by $k_{0}(1+\alpha_{E} \Delta T)$. The relative frequency shift is thus independent of the mode number $n$ and evaluates to
\begin{equation} \label{eq:cteT}
\frac{\delta \omega_{n}}{\omega_{n}} =  \frac{1}{2}  \frac{\delta k_{0}}{k_{0}} = \frac{1}{2}  \alpha_{E} \Delta T
\end{equation}

\subsection{Arbitrary temperature profile}

We now allow any temperature profile, but work in the limit of small perturbations to the stiffness of the cantilever : $\alpha_{E} \theta(x) \ll 1$. We introduce the operator $\mathcal{L}= \partial^{2}/\partial x^{2} [(1+\alpha_{E} \theta(x))\partial^{2}/\partial x^{2} ]$, which is still self-adjoint on space $\mathcal{D}$. The eigenvectors $\phi_{n}$ associated to the eigenvalues $\lambda_{n}$ of this operator form an orthonormal basis of $\mathcal{D}$. They are defined by:
\begin{equation}
\mathcal{L} \phi_{n} = \lambda_{n} \phi_{n} 
\end{equation}
Let us expand this relation to first order in $\alpha_{E}$ with $\mathcal{L}=\mathcal{L}_{0}+\alpha_{E}  \mathcal{L}_{1}$, $\phi_{n}=\phi_{n}^{0}+\alpha_{E} \phi_{n}^{1}$ and $\lambda_{n}=\lambda_{n}^{0}+\alpha_{E} \lambda_{n}^{1}$:
\begin{equation}
\mathcal{L}_{0} \phi_{n}^{1} + \mathcal{L}_{1} \phi_{n}^{0} = \lambda_{n}^{0} \phi_{n}^{1}+\lambda_{n}^{1} \phi_{n}^{0}
\end{equation}
Projecting this last equation on $\phi_{n}^{0}$ and using the property of $\mathcal{L}_{0}$ being self-adjoint, we find
\begin{equation}
\lambda_{n}^{1} = \phi_{n}^{0} \odot \mathcal{L}_{1} \phi_{n}^{0}
\end{equation}
Using the explicit expression $\mathcal{L}_{1} =  \partial^{2}/\partial x^{2} [\theta(x)\partial^{2}/\partial x^{2}]$ and some simple calculations, it is straightforward to show that
\begin{equation}
\lambda_{n}^{1} = \int_{0}^{1} \d x \, \theta(x) \left(\frac{\d^{2} \phi_{n}^{0}}{\d x^{2}}\right)^{2}
\end{equation}
For any given temperature profile $\theta(x)$, it is therefore easy to compute the induced shift of the spatial eigenvalue. We can then use the dispersion relation (\ref{eq:dispertion}) to deduce the shift in resonance frequency:
\begin{align}
\frac{\delta \omega_{n}}{\omega_{n}} &= \frac{1}{2} \frac{\alpha_{E}\lambda_{n}^{1}}{\lambda_{n}^{0}} \\
&= \frac{1}{2} \alpha_{E} \frac{1}{\alpha_{n}^{4}} \int_{0}^{1} \d x \, \theta(x) \left(\frac{\d^{2} \phi_{n}^{0}}{\d x^{2}}\right)^{2} 
\label{eq:frequency-shift}
\end{align}
The frequency shift is thus mode dependent: the softening ($\alpha_{E}<0$) of the cantilever due to the temperature rise is weighted by the square of the local curvature, that is where the cantilever is most bended. Note that if we choose a constant temperature profile $\theta = \Delta T$, we recover the mode independent frequency shift of equation (\ref{eq:cteT}).

It is important to notice that the formal expression (\ref{eq:frequency-shift}) of the frequency shift is independent of the dissipation mechanism, whether viscous (from the surrounding fluid) or viscoelastic (internal). The presence of air around the cantilever only influences the frequency shift due to inertial effects (variation of $m_{\eff}$ with $T$), and through its thermal conductivity which alters the temperature profile.

\subsection{Application: linear temperature profile}

Let us consider the case of a linear temperature profile $\theta(x) = x \Delta T $, where $\Delta T$ is the temperature increase at the free end of the cantilever. We report in table \ref{table:coefdf} the expected coefficients $\kappa_{n}$ linking the relative frequency shift of this temperature profile to the relative frequency shift of a cantilever with uniform temperature $\Delta T$. This coefficient is close to $1/2$ for large mode numbers which present a strong curvature all cantilever long: we observe the effect of the mean temperature $\Delta T/2$. For low mode numbers however (especially mode 1), this coefficient is smaller than $1/2$: curvature is larger close to the clamped end which is at lower temperature than average, thus the frequency shift is not that large.

\begin{table}[htbp]
\begin{center}
\begin{tabular}{|c|c|c|c|c|c|c|}
\hline
Mode number & 1 & 2 & 3 & 4 & 5 & 6 \\
\hline
$\kappa_{n}=\dfrac{\delta \omega_{n}}{\omega_{n}} \dfrac{2}{\alpha_{E}\Delta T}$ & 0.193 & 0.406 & 0.476 & 0.483 & 0.490 & 0.493 \\
\hline
\end{tabular}
\end{center}
\caption{Relative frequency shift for a cantilever with a linear temperature profile normalized to the relative frequency shift of a cantilever with the same maximum temperature but a flat profile.}
\label{table:coefdf}
\end{table}

\section{Temperature profile of a cantilever in vacuum heated at its extremity\label{section:discussion}}

Let us first consider the case of the measurements in vacuum. In this case, the main dissipation process for the heat due to the absorption of light by silicon is thermal conduction along the cantilever. The stationary temperature profile can be estimated using Fourier law for heat transfer:
\begin{equation} \label{eq:Fourier}
J = \frac{a I}{W H} = \lambda(T(x)) \frac{1}{L} \frac{\d T}{\d x} 
\end{equation}
where $J$ is the absolute value of the heat flux (in $\SI{}{W/m^{2}}$), $a$ is the fraction of light absorbed by the cantilever, and $\lambda$ the thermal conductivity of silicon. We assume here that the heat flux is homogeneous across the cantilever cross section $W H$. The temperature profile can be computed from this equation. 

\subsection{Low intensities: linear approximation}

\begin{figure}[htbp]
\begin{center}
\includegraphics{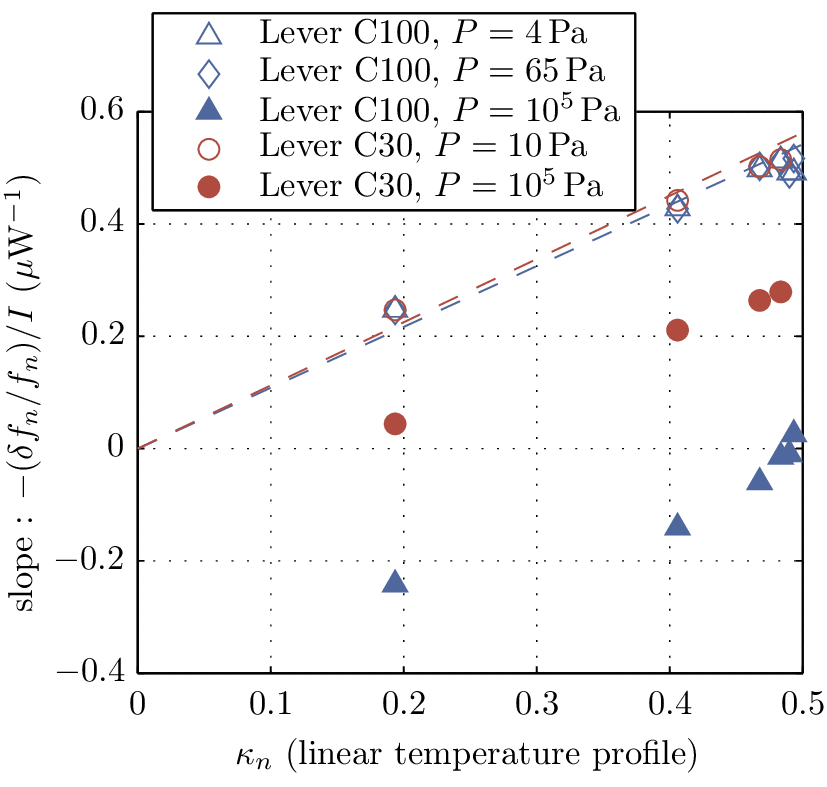}
\caption{(Color online) --- Slope of the relative frequency shift versus light power as a function of coefficient $\kappa_{n}$ (expected mode dependence for a linear temperature profile). The proportionality (dashed lines) is reasonable for both cantilevers in vacuum. At atmospheric pressure, a global vertical offset (mode independent contribution) can be observed. }
\label{fig:dfslope}
\end{center}
\end{figure}

For low light intensities $I$, we neglect the temperature dependence of $\lambda$, the temperature profile is then trivially linear:
\begin{equation}
\theta(x) = \frac{L a I}{W H \lambda_{0}}  x 
\end{equation}
In such a case, the dependence of the frequency shift on mode number $n$ should be modeled by the coefficients $\kappa_{n}$ of table \ref{table:coefdf}, computed for a linear temperature profile :
\begin{equation} \label{dfvsI}
\frac{\delta \omega_{n}}{\omega_{n}} = \kappa_{n} \frac{\alpha_{E}\Delta T}{2} = \kappa_{n} \frac{\alpha_{E}L a}{2 W H \lambda_{0}} I 
\end{equation}
To test this dependence, we perform linear fits of the curves of figures \ref{fig:dfC100} and \ref{fig:dfC30} (considering only data below $\SI{6}{mW}$ for cantilever C100 to avoid non linearities). The slope of these fits, $\delta f_{n}/f_{n}I$, are plotted against $\kappa_{n}$ in figure~\ref{fig:dfslope}. For both cantilevers in vacuum, the mode dependence is correctly captured by our analysis. The prefactor is very similar for both cantilevers: though their geometry is different, they share very similar length $L$, cross section  $WH$, and coefficient of absorption of light by the cantilever $a$.

\begin{figure}[htbp]
\begin{center}
\includegraphics{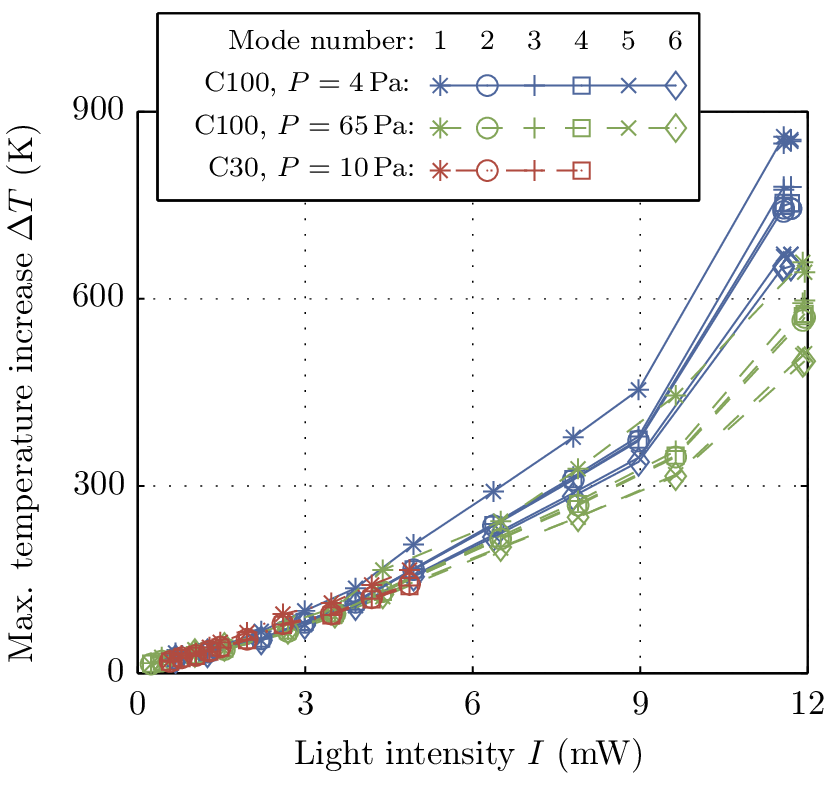}
\caption{(Color online) --- Temperature increase at the free end of the cantilever estimated from the frequency shift in vacuum, assuming a linear temperature profile. This approximation tends to be false (underestimated) at higher power where nonlinearities are higher, and the estimation of $\Delta T$ for the different modes tends to be more disperse. We see however that large temperature gradients are already present for a few mW illumination.}
\label{fig:DTfromdf}
\end{center}
\end{figure}

Using Eq.~(\ref{dfvsI}), we can also estimate the maximum temperature reached at the cantilever free end from the experimental frequency shift. As seen in figure \ref{fig:DTfromdf}, this estimation is coherent for the different modes. The temperature rise is significant: around $\SI{100}{K}$ for a $\SI{3}{mW}$ illumination, and up to $\SI{800}{K}$ for $\SI{12}{mW}$ ! For the largest light intensities, the curve is not linear anymore: the dependence of $\lambda$ on temperature cannot be neglected.

\subsection{High intensities: non-linear approach}

\begin{figure}[htbp]
\begin{center}
\includegraphics{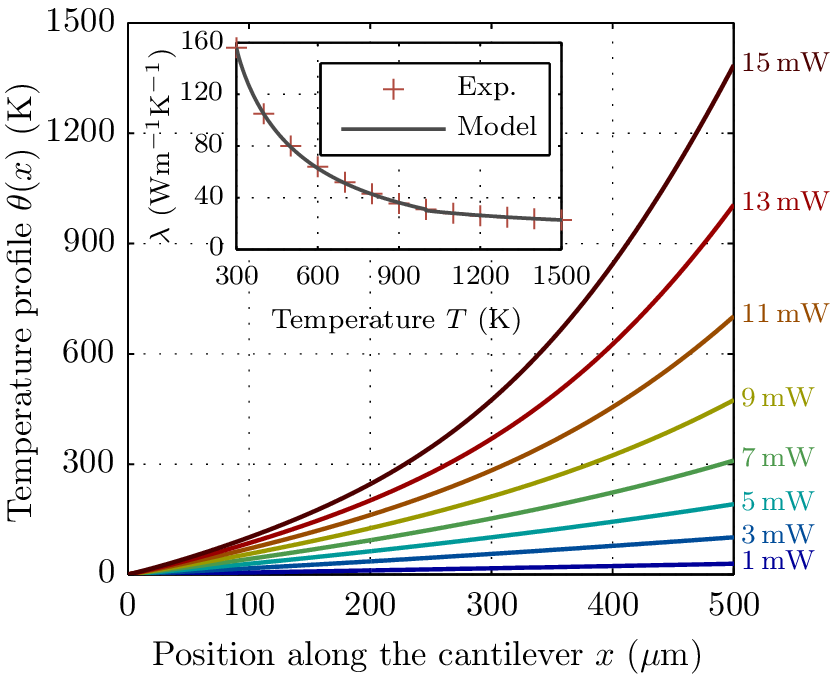}
\caption{(Color online) --- Theoretical temperature profiles computed for a cantilever with a cross section $WH=\SI{71}{\micro m^{2}}$ and an absorption of light $a=0.6$, taking into account the dependence of thermal conductivity of silicon $\lambda$ on temperature. The only thermal transfer process taken into account here is conduction along the cantilever. The profile is clearly nonlinear for light intensities above $\SI{6}{mW}$. Inset: dependence of the thermal conductivity of silicon $\lambda$ on temperature $T$~\cite{Glassbrenner-1964,Prakash-1978}. The variation of $\lambda$ is significant and should be taken into account.}
\label{fig:TxC100}
\end{center}
\end{figure}

We plot in the inset of figure \ref{fig:TxC100} the behavior of $\lambda(T)$ from references~\cite{Glassbrenner-1964,Prakash-1978} (experimental data and model). We can use these data to compute the temperature profile as follows: defining
\begin{equation}
\Lambda (T) = \int_{T_{0}}^{T} \lambda(T') \d T',
\end{equation}
we have by integrating equation~(\ref{eq:Fourier}) over $x$ that
$JLx = \Lambda(T(x))$, from which the temperature profile is given by
\begin{equation}
\theta(x) = \Lambda^{-1} (JLx) - T_0.
\end{equation}

\begin{figure}[htbp]
\begin{center}
\includegraphics{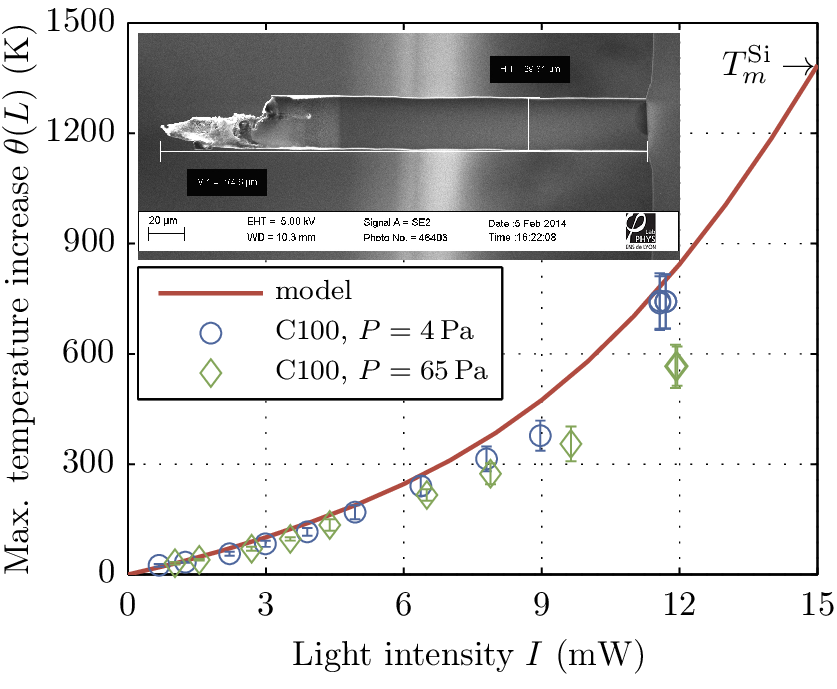}
\caption{(Color online) --- Maximum temperature reached at the end of cantilever C100 as a function of light intensity: model (plain line) and experimental data from figure \ref{fig:DTfromdf} (mean temperature extracted from the 6 measured modes, the uncertainty corresponds to the standard deviation between those 6 points). Inset : SEM Image of a C30 cantilever heated by a laser power of $I=\SI{25}{mW}$: the temperature raised above $T_{m}^{\mathrm{Si}}=\SI{1410}{\degree C}$, leading to the melting of silicon. The initial length $L=\SI{500}{\micro m}$ has been shorten to $\SI{275}{\micro m}$ when moving the laser focus towards the base.}
\label{fig:TmaxC100}
\end{center}
\end{figure}

From the data in the inset of figure \ref{fig:TxC100}, we can estimate numerically $\Lambda (T)$ and thus what the temperature profiles are for various light intensities $I$. We plot these estimations in figure \ref{fig:TxC100} for cantilever C100, assuming an absorption coefficient $a=0.6$. A more accurate estimate of the temperature profile would require to know the temperature dependence of $a$; here we consider for simplicity a typical, temperature-independent value of $a$.
The nonlinear shape of the temperature profiles are clearly visible for light powers above $\SI{6}{mW}$. We report the maximum temperature expected from these non linear profiles (at the free end of the cantilever) in figure \ref{fig:TmaxC100}. The agreement with the experimental data is good, especially as the measurement points are relying of a linear temperature profile analysis. Note that in principle, non-linear temperature profiles should also be considered in the analysis of the frequency shifts. However, since the precise, temperature-dependent value of the light absorption coefficient is unknown, the accuracy of the determination method would be difficult to assess, and we thus stick to the analysis based on a linear temperature profile.

Since the melting point of silicon is $T_{m}^{\mathrm{Si}}=\SI{1410}{\degree C}$, we see that this analysis predicts dramatic consequences for the cantilever for light intensities higher than $\SI{15}{mW}$ ! And indeed, for the highest intensities we have probed in vacuum, some cantilever have been molten at their extremity. We present for example in figure \ref{fig:TmaxC100} a SEM image of a C30 cantilever, that could be shorten from its initial length $L=\SI{500}{\micro m}$ to only $\SI{275}{\micro m}$ using $I=\SI{25}{mW}$.

\begin{figure}[htbp]
\begin{center}
\includegraphics{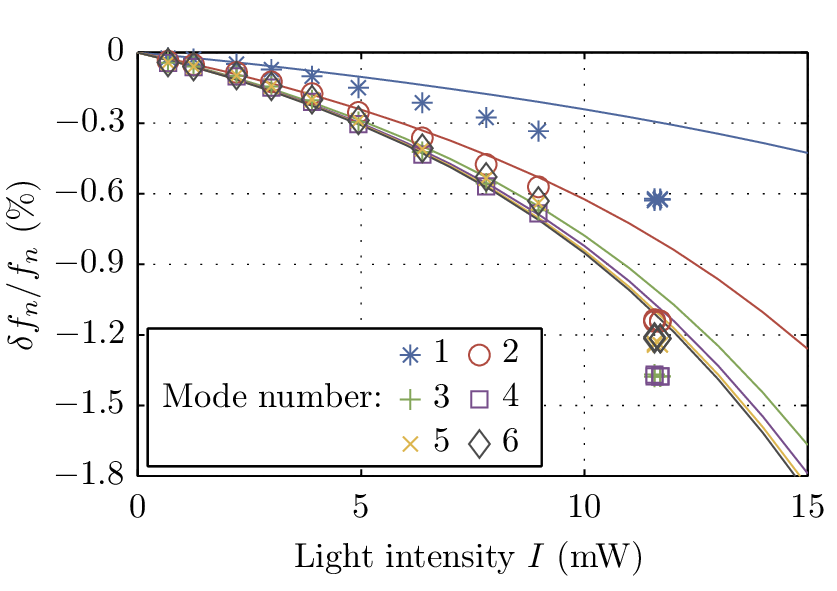}
\caption{(Color online) --- Theoretical (plain lines) and experimental (symbols) relative frequency shift $\delta f_{n}/f_{n}$ for mode $n=1$ to $6$ as a function of incident light power $I$ on the lever. The theoretical curves have been computed for a cantilever with a cross section $WH=\SI{71}{\micro m^{2}}$ in vacuum, with an absorption of light of $a=0.6$, taking into account the dependence of thermal conductivity of silicon $\lambda$ on temperature. The experimental data is that of figure \ref{fig:dfC100} in vacuum at $\SI{4}{Pa}$.}
\label{fig:dfthC100}
\end{center}
\end{figure}

Using the temperature profiles of figure \ref{fig:TxC100}, we can estimate the frequency shifts from equation (\ref{eq:frequency-shift}). The results are plotted in figure \ref{fig:dfthC100}. The curves are qualitatively very similar to the experimental measurements of figure \ref{fig:dfC100}. A quantitative agreement cannot be achieved anyway, and we can point a least 3 levels of approximation :
\begin{enumerate}
\item the temperature dependence of the Young's modulus of silicon would have to be better known in the temperature range of interest.
\item the coefficient of absorption $a$ depends on temperature. 
\item the temperature is not uniform across the cantilever cross section close to the free end where the laser is focused.
\end{enumerate}

The general framework presented here is thus appropriate to describe the behavior of cantilevers in vacuum heated by light absorption at their free extremity. Through the proposed analysis, the frequency shifts prove to be a robust way of estimating the temperature.

\section{Inertial effects of the surrounding fluid in air\label{section:air}}

In air, the observed frequency shift depends strongly on the cantilever shape, and can even be opposite to the expected trend: the resonance frequency increases with laser power for cantilever C100. However, additional effects have to be taken into account due to the presence of the atmosphere: the temperature rise of the cantilever heats the surrounding air, increasing its viscosity and lowering its density. The latter effect leads to a decrease of the inertial effects due to the air moving along with the cantilever, hence help raising the resonance frequency~\cite{Mertens-2003}. This effect is more pronounced on larger cantilevers since the volume of air implied is higher.

We did not conduct a systematic study of the interaction between this effect and a non uniform temperature profile, but a first estimation of the effect of the temperature rise in air can be done with the Sader model~\cite{Sader-1998}. Indeed, from equation \ref{eq:dispertion} for a uniform temperature $T_{0}$, we can compute for every cantilever the resonance frequency shift:
\begin{equation}
\frac{\delta \omega_{n}}{\omega_{n}} =  \frac{1}{2}  \frac{\delta k_{0}}{k_{0}} - \frac{1}{2}  \frac{\delta m_{\eff}}{m_{\eff}}
\end{equation}
The first term accounts for the temperature dependence of the Young's modulus of Silicon, while the second term accounts for the variation of the effective mass of the oscillator. Let us explicit $m_{\eff}$:
\begin{equation}
m_{\eff}(T_{0})=m+\frac{\pi}{4}L W^{2} \rho(T_{0}) \Gamma_{r}(\omega,\nu(T_{0}))
\end{equation}
In this equation, we neglect again the temperature dependence of the geometry ($L$ and $W$ supposed constant), but consider explicitly the variation of the air density $\rho$ and kinematic viscosity $\nu$ with $T_{0}$~\cite{Sutherland-1893}:
\begin{align}
\rho (T_{0}) &= \rho_{\star}\frac{T_{\star}}{T_{0}} \\
\nu (T_{0}) & =\nu_{\star}\frac{T_{\star}+S}{T_{0}+S}\left(\frac{T_{0}}{T_{\star}}\right)^{5/2}
\end{align}
with $\rho_{\star}=\SI{1.29}{kg/m^{3}}$, $T_{\star}=\SI{273}{K}$, $S={110.4}{K}$ and $\nu_{\star}=\SI{1.32e-5}{m^{2}/s}$. From these equations and the explicit expression of the hydrodynamic function $\Gamma(\omega,\nu)$ in reference~\cite{Sader-1998}, we can compute $\delta m_{\eff}/m_{\eff}$ as a function of a uniform temperature $T_{0}$.

\begin{figure}[htbp]
\begin{center}
\includegraphics{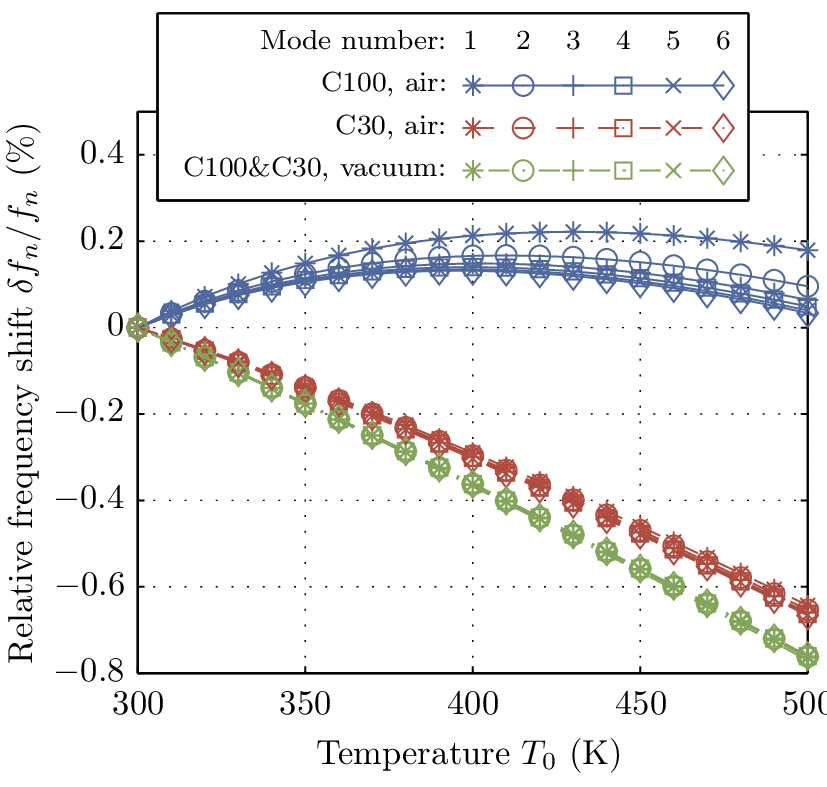}
\caption{(Color online) --- Theoretical relative frequency shift $\delta f_{n}/f_{n}$ versus uniform temperature $T$ of cantilever and air according to the Sader model. The shift is the sum of two opposite effects when temperature rises: the softening of the cantilever (only effect in vacuum), and the decreasing of the inertial effects due to the surrounding atmosphere. The latter effect is dominant for the wider cantilever at low $T$. For a uniform temperature, the mode dependency is weak.}
\label{fig:dfthSader}
\end{center}
\end{figure}

We plot in figure \ref{fig:dfthSader} the frequency shift expected for a silicon cantilever as a function of $T_{0}$. We observe that when $T_{0}$ increases, the inertial effects due to the decreasing density of air are opposite to the softening of the Young's modulus of silicon. For the wider cantilever, the former effect is dominant, whereas for the narrower one, it is just a correction to the latter. Though the temperature profile is not considered in this model, a similarity with the experimental data of figure \ref{fig:dfC100} can be seen : the frequency shifts of intermediate modes present a maximum and then decay, when inertial effects due to the fluid vanish at high temperature.

If the temperature profile was to be considered, dissipation of heat through air would lead to a temperature rise located at the end of the cantilever only. The inertial effects would thus be more important on mode 1, for which the highest deflection are close to the end. At the same time, the base of the cantilever, where most of the deformation occurs, will not see the temperature rise, thus softening of the cantilever is negligible. And indeed, the maximum for mode 1 is not seen in the experimental data of figure \ref{fig:dfC100}, the inertial effects being always dominant.

As a final note, if we focus on the frequency shift due to the inertial effects only in this model, we checked that it depends only weakly on the mode number. The slope of the frequency shift versus light power should therefore just be translated by a mode independent contribution in air with respect to vacuum, as observed on the experimental data of figure \ref{fig:dfslope}.

\section{Conclusion}

To measure the temperature profile of the cantilever, we propose to use the frequency shifts triggered by the softening of silicon upon heating. The proposed model adequately describe the experimental data in vacuum below $\SI{200}{\degree C}$, and could be quantitative at higher temperature provided the Young's modulus of silicon is better characterized in the adequate temperature range. Our approach is robust for all flexural modes of the cantilever, and relates the frequency shift to the average of the temperature profile weighted by the local curvature. Coefficients to apply to modes 1 to 6 are given in table \ref{table:coefdf} for direct application to linear temperature profiles. As for observations in air, a qualitative description of the measured frequency shift has been proposed through the dependence of inertial effects due to the surrounding fluid in its temperature.

As demonstrated in this article, strong heating of a raw silicon AFM cantilever in vacuum can be achieved using light absorption of a laser focused at the free end of the cantilever. The temperature rise depends on the length and cross section, but can be larger than a thousand $\SI{}{\degree C}$ for a $\SI{500}{\micro m}$ long cantilever with little more than $\SI{10}{mW}$ at $\SI{532}{nm}$. Such large laser intensities are not common in standard AFM operation, but the usual $\SI{1}{mW}$ is already enough to produce a $\SI{30}{K}$ increase, which should be considered for temperature sensitive samples. Reflex coated cantilevers, or shorter ones, will obviously reduce this issue. On the other hand, if one wants to heat locally a sample (to trigger phase changes or modify other properties), high temperature of the AFM tip can be reached through a simple external laser focused on the cantilever free end. In this case, the current framework provides an easy procedure to estimate the temperature of the AFM probe.

\begin{acknowledgments}
We thank F. Vittoz and F. Ropars for technical support, A. Petrosyan, S. Ciliberto for stimulating discussions. This work has been supported by the ERC project \emph{OutEFLUCOP} and the ANR project \emph{HiResAFM} (ANR-11-JS04-012-01) of the Agence Nationale de la Recherche in France.
\end{acknowledgments}

\bibliography{freqshift.bib}

\end{document}